# Ambiguity Resolution with Human Feedback for Code Writing Tasks

**Aditey NANDAN[a] & Viraj KUMAR[b*]**
[a]*Mathematics and Computing, Indian Institute of Science, India*
[b]*Kotak-IISc AI-ML Centre, Indian Institute of Science, India*
*viraj@iisc.ac.in

**Abstract:** Specifications for code writing tasks are usually expressed in natural language and may be ambiguous. Programmers must therefore develop the ability to recognize ambiguities in task specifications and resolve them by asking clarifying questions. We present and evaluate a prototype system, based on a novel technique (Ambiguity Resolution with Human Feedback: ARHF), that (1) suggests specific inputs on which a given task specification may be ambiguous, (2) seeks limited human feedback about the code's desired behavior on those inputs, and (3) attempts to generate code that resolves these ambiguities. We evaluate the efficacy of our prototype, and we discuss the implications of such assistive systems on Computer Science education.

## 1. Introduction

Real-world specifications for code writing tasks are often expressed in natural language, and such specifications can be ambiguous (Gervasi, et al., 2019; Jinwala & Shah, 2015). A failure to resolve such ambiguities early in the software development lifecycle can lead to costly errors (Fernández et al., 2017). Therefore, existing computing curricula such as CS2023 (ACM/IEEE-CS/AAAI Joint Task Force, 2023) emphasize the importance of requirements elicitation (pg. 242) and regard the ability of students to assess the "lack of ambiguity" in technical documents as a core learning outcome (pg. 280). The CS2023 curriculum further notes that, in the context of networking and communication, "Generative AI technologies ... can verify whether networking requirements are ambiguous or complete" (pg. 450). Since students have easy access to powerful Generative AI tools, and since these tools are increasingly being used in professional software development contexts, Becker et al. (2023) have highlighted an urgent need to review educational practices in computing. In this spirit, we ask the following question: Can assistive systems help programmers identify and resolve ambiguities in task specifications? If so, we are interested in understanding the impact of such systems on computing education. In this paper, we limit our attention to code writing in the context of introductory programming courses (CS1).

Figure 1 shows a typical CS1 code-writing task with a specification that is ambiguous because it is incomplete: the purpose statement (or docstring) on Line 2 and the example (or doctest) on Lines 3 and 4 fail to specify the code's expected behavior on certain inputs. As noted by Schneider (1978), such ambiguous specifications can be interpreted in multiple "reasonable" ways that are functionally inequivalent. Before writing any code, a good programmer would therefore ask clarifying questions to distinguish between these inequivalent interpretations. Instead of seeking clarity, modern code-generation tools such as GitHub Copilot often make one or more assumptions. For instance, the code in Figure 1 makes two seemingly "reasonable" assumptions: the function should return -1 if the input string *s* has no digits (Assumption 1), and it should return the index of the first occurrence if the smallest digit occurs multiple times in *s* (Assumption 2). However, we think it is "unreasonable" for the code to assume that all letters in *s* are digits (Assumption 3). Our goal is not to improve such code-generation tools. Instead, we seek to proactively explore whether it is possible for assistive systems to help programmers recognize that a given task supports multiple interpretations, and to distinguish between these interpretations by suggesting clarifying questions. If so, we hope to identify the skills that students will need to develop if such tools become widely available.

```
1  def min_index(s: str) -> int:
2      """Find the index of the smallest digit ('0' to '9') in s.
3      >>> min_index('2025')
4      1
5      """
6      min_digit = '9'
       min_index = -1
       for i, c in enumerate(s):
           if c < min_digit:
               min_digit = c
               min_index = i
       return min_index
```

GitHub Copilot's suggested solution (Line 6 onwards)

Clarifying Questions
1. What if the string has no digits?
2. What if the smallest digit appears more than once?

Assumption 1: If no digits, return -1.
Assumption 3: No non-digits.
Assumption 2: If the smallest digit appears more than once, return the *first* index.

Ambiguity Resolution
Some inputs suggested by our system:
1. `''` # empty string
2. `'00000000'`

*Figure 1.* A solution generated by GitHub Copilot solution (Line 6 onwards) for an ambiguous code-writing task (problem P1 in this study, Lines 2 to 5). The generated code resolves these ambiguities by making assumptions. A good programmer would recognize these ambiguities and ask at least the two clarifying questions shown. Instead, our system generates inputs and seeks a limited form of human feedback to resolve these ambiguities.

To this end, we develop a prototype assistive system. We initially attempted to use language models such as GPT-4o to generate general clarifying questions from ambiguous task specifications (such as in Figure 1), but the generated questions were of poor quality. We thus propose a technique, Ambiguity Resolution with Human Feedback (ARHF), whose novelty stems from the insight that code generation models can be prompted to:
1. Use an ambiguous task specification to generate *specific* inputs on which the code's desired behavior *may* be ambiguous, and
2. Use human feedback that clarifies the code's desired behavior on these specific inputs to produce code that handles *general* inputs correctly.

We have developed a prototype system that uses ARHF, and we have evaluated it on six CS1-level code-writing tasks in Python. Our research questions are:
- **RQ1**: How effectively can our ARHF system generate inputs that distinguish between all reasonable interpretations of an ambiguous task specification?
- **RQ2**: How effectively can our ARHF system use human feedback to generate accurate code given an ambiguous task specification?

We find that although ARHF is effective at generating inputs that distinguish the target interpretation from other reasonable interpretations, it is sometimes unable to generate the implementation for this interpretation.

## 2. Related Work

There has been substantial interest in the Computer Science education research community to rethink pedagogy, assessments, and learning outcomes now that students have easy access to powerful Generative AI tools, particularly since these tools are being integrated into professional workflows (Prather et al., 2023, Raman & Kumar, 2022). For instance, code-writing tasks in introductory programming are typically simple and well-specified, and large language models trained to generate code (henceforth: Code LLMs) can solve such tasks reliably (Prather et al., 2023).

Almost five decades ago, Schneider (1978) called for code-writing tasks where certain details are deliberately omitted from the task specification, thereby allowing multiple interpretations. Such code-writing tasks require students to assume a "reasonable" target interpretation. For such a task, Code LLMs such as GitHub Copilot can often generate code corresponding to a reasonable interpretation, as illustrated in Figure 1. Recent research has therefore explored a variation of Schneider's proposal – Probeable Problems (Pawagi & Kumar, 2024; Denny et al., 2025). A Probeable Problem is once again a code-writing task where the specification can be interpreted in multiple ways because it lacks certain details, but where students can issue 'probes' to clarify the target interpretation before writing code. Each probe is a test input, and the behavior of the target interpretation is revealed by executing the (hidden) target implementation on that input.

In contrast, we propose an assistive system that is similar to interactive test-driven code generation (Lahiri et al., 2022) where both approaches: (1) assume that the given task specification S may be ambiguous (i.e., it may be consistent with multiple reasonable interpretations), (2) assume that the programmer cannot access a target implementation $X$ but does have a target interpretation of $S$ in mind, (3) use a Code LLM (which may make assumptions) to generate a candidate implementation $C$ from specification $S$, (4) generate a set of test inputs $T$ that may reveal differences between multiple interpretations of $S$, (5) execute candidate implementation $C$ on each input in $T$, and (6) use the programmer's understanding of the target interpretation of $S$ to obtain feedback on these executions and hone in on the target implementation $X$. The key difference in these two approaches is the nature of the human feedback: whereas Lahiri et al. (2022) merely ask for "lightweight" Boolean feedback (i.e., whether or not implementation $C$ has the desired behavior on each input in $T$), we expect programmers to be capable of accurately specifying the behavior of $X$ whenever this differs from the behavior of $C$ on an input in $T$. Hebbar et al. (2025) call this skill reactive task comprehension. We name our system as per our approach: Ambiguity Resolution with Human Feedback (ARHF).

## 3. Methods

*3.1 ARHF System Architecture*

The input to ARHF is a (potentially ambiguous) specification $S$. For this study, we assume that $S$ consists of a Python function *signature* (the function name as well as the name and type-hint for each argument), a *docstring* explaining the function's purpose, and at least one *doctest* (an example showing the desired output as per the target interpretation on some test input). For the problem shown in Figure 1, the specification $S$ (Lines 1 to 5) is ambiguous because it is consistent with many reasonable interpretations, two of which are:

- $I_1$: Find the *first* index of the smallest digit ('0' to '9') in *s*. *If no such digit exists, return* −*1.*
- $I_2$: Find the *first* index of the smallest digit ('0' to '9') in *s*. *If no such digit exists, return len*(*s*).

ARHF has three components: Initial Code Generation (which generates a candidate implementation $C$ from the given implementation $S$), Test Input Generation (which generates test inputs $T$ from $S$ and $C$, and Code Correction (which attempts to correct implementation $C$ to the target implementation $X$ using human feedback on $C$'s output for each input in $T$). All three components may use different Code LLMs. However, our prototype system uses a common Code LLM: Qwen2.5-Coder-32B-Instruct (Hui et al., 2024). The detailed prompt for each component is provided in Table 1.

We illustrate how these components work for the problem in Figure 1. The Initial Code Generation component uses the prompt in Table 1, and for this specification $S$, the Code LLM frequently generates a candidate implementation $C_1$ corresponding to interpretation $I_1$. Note that $C_1$ passes the given doctest (lines 3 to 4 in Figure 1). However, if the candidate implementation fails the provided doctest, this failure is recorded for later.

Next, the Test Input Generation component generates a set $T$ of test inputs in two ways. First, it prompts the Code LLM as shown in Table 1. Second, it invokes CrossHair (Schanely, 2017), a symbolic execution tool, on the candidate implementation to generate additional test inputs. We find that symbolic execution helps our system generate test inputs that specifically target edge cases. Each test input in this combined set $T$ is then executed on the candidate implementation, and our system displays each input-output pair as a doctest. Three such doctests are shown in Figure 2 for the **min_index** problem P1. If interpretation $I_1$ is in fact the target interpretation, then the programmer can confirm that the candidate implementation $C_1$ passes all doctests. The ARHF front-end pre-selects the "Accept" option, as shown in Figure 2 (left). In this case, there are no failing doctests, and the candidate implementation $C_1$ is revealed as the desired code. On the other hand, if the target interpretation is $I_2$, the candidate implementation is buggy. In this case, the programmer should select "Reject" for the first two doctests and should specify the correct output: 0 for the empty input string, and 5 for the input string 'abcde' as shown in Figure 2 (right).

Table 1. *Detailed prompts for each component of ARHF*

| Component | Detailed Prompt |
|---|---|
| Initial Code Generation | Consider the following Python function signature, docstring, and doctests:<br>{function_signature}<br>{docstring + doctests}<br>Write ONLY the code for the function body. Ensure that the resulting function passes ALL doctests. |
| Test Input Generation | You are an expert at designing test inputs to discover potential ambiguities in the docstring of a Python function. Consider the following Python function signature and docstring, which may include doctests:<br>{function_signature}<br>{docstring + doctests}<br>Generate a list `test_inputs`, where each list item is a tuple corresponding to the function's arguments. The inputs you generate must explore the space of legal inputs thoroughly enough to discover potential ambiguities. DO NOT generate the function's expected outputs for the inputs in `test_inputs`.<br>test_inputs = [ |
| Code Correction | Although the following Python function passes the given doctests, it is BUGGY because the docstring is ambiguous i.e., it does not fully clarify the intended purpose:<br>{function_signature}<br>{docstring}<br>{passing_doctests}<br>{buggy_code}<br><br>The above function FAILS the following doctests, which help clarify the function's intended purpose:<br>{failing_doctests}<br><br>Rewrite or modify the function so that it satisfies ALL doctests. Remember that:<br>• Some doctests fail because they correspond to edge cases. If necessary, modify the code to handle these cases separately.<br>• Other doctests fail because the above code incorrectly generalizes the intended purpose. You must infer the intended general purpose from these doctests. |

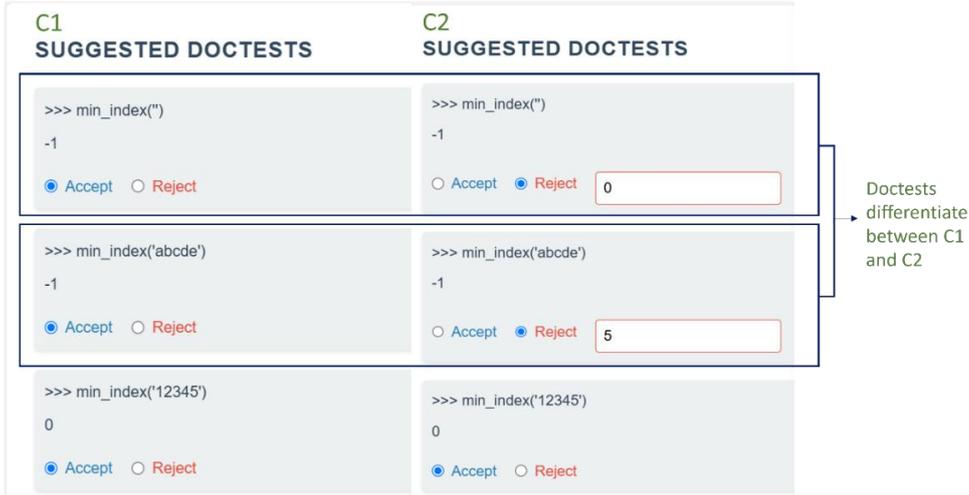

*Figure 2.* The ARHF system's front-end to obtain human feedback for ambiguity resolution. The two highlighted doctests help differentiate between candidate implementations $C_1$ and $C_2$ for the **min_index** problem (P1 in this study)

All failing doctests are sent to the Code Correction component along with the buggy candidate implementation. For this problem, the Code LLM is expected to inductively deduce from these failing doctests that *len*(*s*) is the correct general expression to return when string *s* has no digits. Our system executes the corrected implementation against all doctests. If one or more doctests fail, the Code Correction component can be re-executed until the generated function passes all the corrected doctests or a threshold number of attempts is exceeded. In the latter case, our system presents the "best" candidate implementation to the programmer. This is defined by the function which passes the maximum number of doctests (and the most recently generated function in case of a tie). In this study, we do not perform re-execution. The flowchart for this simplified execution of our system is shown in Figure 3.

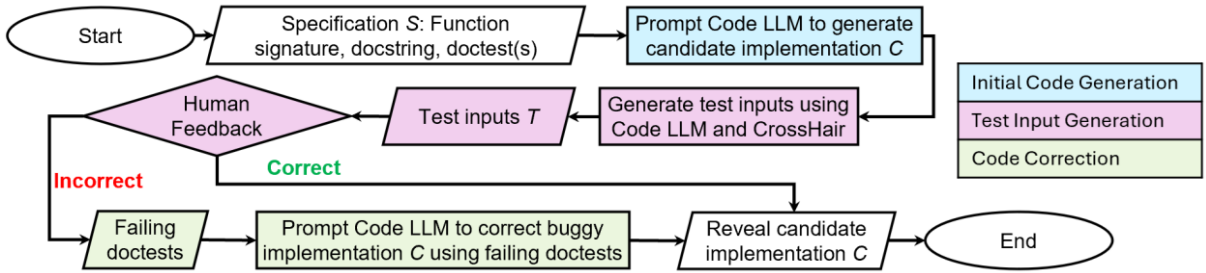

*Figure 3. Flowchart describing the core operation of our ARHF system.*

### 3.2 Components and Implementation

Our ARHF system is implemented as a web based, locally run application. The front-end is built using HTML, CSS and JS, while the backend is implemented using Python (Flask). We used the Hugging Face Inference API to run the Code LLM Qwen2.5-Coder-32B-Instruct (Hui et al., 2024). We use the default settings for the Code LLM: temperature = 0.7, top_p = 0.8, and top_k = 20 (Link to the Code LLM: https://huggingface.co/Qwen/Qwen2.5-32B-Instruct/blob/main/generation_config.json).

### 3.3 Code-writing Problems

We evaluate our system using two types of problems. First, we create three problems, each with multiple interpretations (Table 2). Treating each of these interpretations as the target, we attempt to use ARHF to generate the desired implementation (25 times for each interpretation).

Table 2. *Problems P1, P2, and P3 created for this study, each with multiple reasonable interpretations, the implementations corresponding to these interpretations are available at: https://onlinegdb.com/hFNmVRe-B*

| Problem | Docstring | Initial Doctest | Interpretations |
|---|---|---|---|
| P1 | Find the index of the smallest digit ('0' to '9') in *s*. | >>> min_index('2025')<br>1 | 4 |
| P2 | Find the minimum frequency. Return None if *data* is empty. | >>> min_freq([1, 2, 1])<br>2 | 3 |
| P3 | Count the number of digits in *n*. | >>> num_digits(123)<br>3 | 2 |

Table 3. *Problems P7, P8 (slightly adapted), and P9 from Denny et al. (2025), each with 16 reasonable interpretations, the implementations for these interpretations are available at: https://onlinegdb.com/WSOq_Xexb*

| Problem | Docstring | Initial Doctest |
|---|---|---|
| P7 | Count the number of integers in data between a and b. | >>> count_between([1, 2, 3], 0, 5)<br>3 |
| P8 | Search data for the smallest even value. | >>> smallest_even([50, 25, 2, 30, 45])<br>[2] |
| P9 | Find the first vowel in s. | >>> first_vowel('apple')<br>'a' |

In addition, we selected three Probeable Problems proposed by Denny et al., 2025, each with 16 reasonable interpretations as shown in Table 3. We lightly modified problem P8 to return a value rather than to print an answer. For these problems, we fixed the target interpretation as the one chosen by Denny et al. (2025), and we attempted to use ARHF to generate the corresponding implementation (100 times each, as shown in Figure 4).

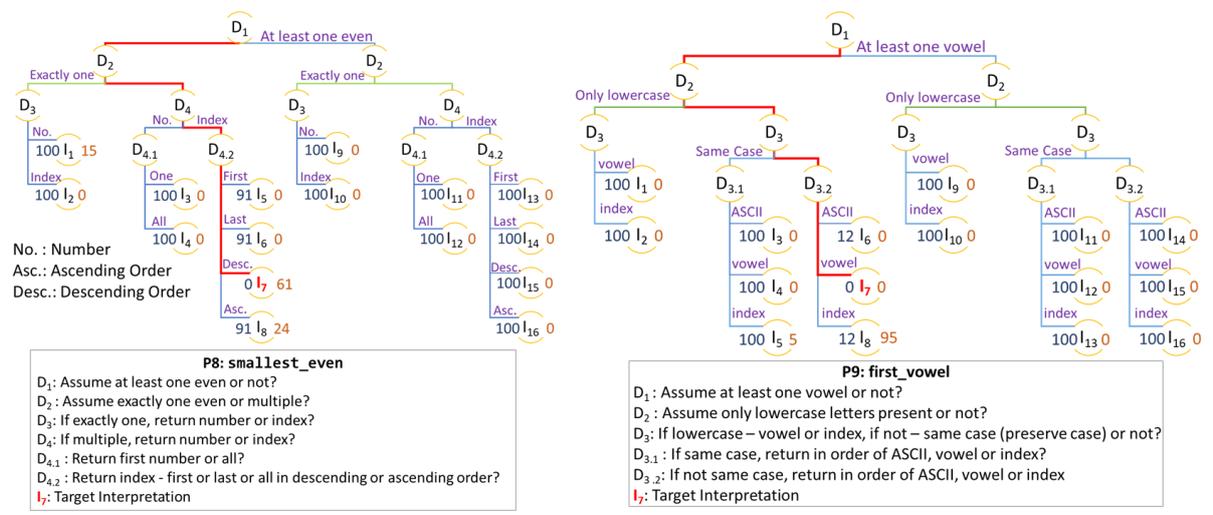

Figure 4. Problems P8 and P9. The leaves of each tree represent the 16 reasonable interpretations while the internal nodes represent decisions that distinguish interpretations. The target interpretation is $I_7$ for both problems, with the decisions leading to that interpretation highlighted as red edges. The leaf node values are explained in the text.

For each problem, we first evaluate the set of test inputs *T* generated by the Test Input Generation component of our system. Specifically, for each implementation *Y* corresponding to an incorrect interpretation, we check whether there is at least one input in *T* on which the output of *Y* disagrees with the output of *X* (the implementation corresponding to the target interpretation). If so, we say that the set of inputs *T* distinguishes the target implementation *X* from the buggy implementation *Y*. We define Input Ambiguity Resolution (IAR) as the fraction of incorrect implementations that are distinguished by the set of inputs *T*.

Finally, for each problem, we check whether the revealed implementation *C* is functionally equivalent to the target implementation *X* and inequivalent to each implementation that corresponds to a non-target interpretation. For this study, we check for equivalence by manual inspection. We use the standard Pass@1 metric (Chen et al., 2021) to estimate the probability that *C* is functionally equivalent to *X*. In addition, we define Code Ambiguity Resolution (CAR) as the fraction of implementations corresponding to non-target interpretations that are functionally inequivalent to *C*.

## 4. Results

*4.1 RQ1: How effectively do ARHF-generated test inputs resolve ambiguities?*

We evaluate the Input Ambiguity Resolution of IAR of our system for the six problems in this study. For problems P1, P2, and P3, we treat each interpretation as the target and execute ARHF 25 times each. For problems P7, P8, and P9, we execute ARHF 100 times for the target interpretation chosen by Denny et al. (2025). Our results (Table 4) demonstrate that the test inputs generated by our system achieve the highest possible IAR (1.0) for all problems with only a few reasonable interpretations (P1, P2, and P3), and they maintain a high IAR (above 0.94) even when problems admit greater ambiguity (P7, P8, and P9).

Table 4. *The Input Ambiguity Resolution (IAR) of our system for the six problems in this study*

| Problem | IAR | Remarks |
|---|---|---|
| P1 | (3 × 25)/(3 × 25) | |
| P2 | (2 × 25)/(2 × 25) | Each interpretation chosen 25 times as the target |
| P3 | (1 × 25)/(1 × 25) | |
| P7 | 1495/(15 × 100) | |
| P8 | 1473/(15 × 100) | Single target interpretation executed 100 times |
| P9 | 1412/(15 × 100) | |

While our system performs well for problems P7 and P8, it performs comparatively poorly for problem P9. To understand why, we analyze the test inputs and code generated by ARHF in relation to the 16 reasonable interpretations for each problem, as shown in Figure 4. The two integers at each leaf respectively show the percentage of ARHF invocations for which (*a*) that interpretation was distinguished from the target interpretation $I_7$ by test inputs (left-side number in dark blue font), and (*b*) the revealed implementation corresponded to that interpretation (right-side number in orange font). For example, the non-target interpretation $I_8$ was correctly distinguished from $I_7$ in 91% of the invocations for P8, but in only 12% of the invocations for P9. In the latter problem, the target interpretation of "first vowel" is: "first vowel in vowel order, ignoring differences in upper-case and lower-case vowels". The test inputs generated by ARHF are often unable to distinguish this from interpretation $I_8$: "first vowel in index order, ignoring differences in upper-case and lower-case vowels". Since the task specification lacks so many crucial details, it is perhaps unsurprising that ARHF cannot generate appropriate test inputs. Nevertheless, the high IAR score indicates that the inputs generated can help eliminate many reasonable interpretations from further consideration.

## 4.2 RQ2: How effectively does ARHF resolve ambiguities in the generated code?

We now examine the ability of ARHF to generate the desired implementation for each of the six ambiguous code-writing problems. As shown in Table 5, ARHF generally achieves high Pass@1 and CAR rates, except for problem P3 (when the target interpretation of "count the number of digits" is $I_2$: "count the unique number of digits), as well as problems P8 and P9. As noted in Section 4.1, the particularly poor performance on problem P9 is unsurprising since the test inputs are insufficiently rich to compensate for the many missing details in the task specification. Thus, in Figure 4, 95% of ARHF invocations reveal an implementation corresponding to interpretation $I_8$ instead of the target interpretation $I_7$. While the Pass@1 rate is zero, the high CAR value (more than 0.91) shows that ARHF successfully rejected most of the unwanted interpretations. While this is promising, we also note from Figure 4 that when the revealed implementation is *inequivalent* to the target implementation, it corresponds to an interpretation that is slightly different from the one desired. In other words, when the implementation generated by ARHF is incorrect, it is likely to be subtly buggy!

Recall that our system achieves an IAR of 1.0 for P3 regardless of which interpretation was chosen as the target (Table 4). Thus, when the target implementation is $I_2$, the doctests created from human feedback on test inputs generated by our system are inconsistent with interpretation $I_1$. Nevertheless, the Code LLM in our system generates an implementation corresponding to interpretation $I_1$ of P3 on all 25 invocations. This suggests an inability of current Code LLMs to generate code that strictly adheres to provided doctests, as noted by Heo et al. (2024).

Table 5. *The Pass@1 rate and Code Ambiguity Resolution (CAR) of our system for the six problems in this study*

| Problem | Target Interpretation | Pass@1 | Code Ambiguity Resolution (CAR) |
|---|---|---|---|
| P1 | $I_1$ | 25/25 | 75/(3 × 25) |
|  | $I_2$ | 25/25 | 75/(3 × 25) |
|  | $I_3$ | 25/25 | 75/(3 × 25) |
|  | $I_4$ | 25/25 | 75/(3 × 25) |
| P2 | $I_1$ | 25/25 | 50/(2 × 25) |
|  | $I_2$ | 23/25 | 48/(2 × 25) |
|  | $I_3$ | 22/25 | 47/(2 × 25) |
| P3 | $I_1$ | 25/25 | 25/(1 × 25) |
|  | $I_2$ | 0/25 | 0/(1 × 25) |
| P7 | $I_7$ | 95/100 | 1489/(15 × 100) |
| P8 | $I_7$ | 61/100 | 1323/(15 × 100) |
| P9 | $I_7$ | 0/100 | 1370/(15 × 100) |

## 5. Limitations and Future Work

We note several limitations of this study and discuss how they can be addressed by future work. First, our study is limited to just six code-writing tasks (specifically, writing a single function), in a single programming language (Python). Future work can explore a richer variety of code-writing tasks. Second, although our results demonstrate that ARHF can be effective at resolving ambiguities, our system requires the programmer to accurately specify the desired outputs on inputs suggested by our system. It will therefore be useful to evaluate whether students, especially novice programmers, can make effective use of a system such as ours. Finally, we reiterate that our system is a prototype that can be improved in several ways. For instance, if the implementation generated by the Code Correction component fails some doctests, we can make additional attempts to generate an implementation that passes these doctests. It is also possible that breakthroughs in Code LLMs will improve their ability to adhere to specified doctests, resulting in a more effective system overall.

## 6. Conclusion

In conclusion, this paper has presented a prototype system that uses off-the-shelf components to effectively resolve ambiguities in code-writing task specifications using a novel strategy: Ambiguity Resolution with Human Feedback (ARHF). We have shown that although ARHF is imperfect when key details are missing in the specification, it is nevertheless helpful in rejecting unwanted interpretations of that specification. Therefore, it is plausible to anticipate that students will soon have easy access to similar systems more capable than our prototype. Thus, our study suggests that learning outcomes associated with identifying ambiguities (ACM/IEEE-CS/AAAI Joint Task Force, 2023) and assessments such as Probeable Problems (Pawagi and Kumar, 2024; Denny et al., 2025) may soon need to be redesigned.

## Acknowledgements

We would like to thank the Kotak-IISc AI-ML Centre for supporting this research.